\def \yskip{\penalty-50\vskip3pt plus 3pt minus 2pt}
\def \reference{\par \yskip \noindent \hangindent .4in \hangafter 1}
\def \abc#1#2#3#4 {\reference#1, {\sl#2}, {\bf#3}, #4}
\def \blank {\lower 5pt\hbox to 0.75in{\hrulefill}}

\def \cm{~\rm{cm}}
\def \s{~\rm{s}}
\def \km{~\rm{km}}

\def \g{~\rm{g}}
\def \AU{~\rm{AU}}
\def \yrs{~\rm{yrs}}
\def \yr{~\rm{yr}}
\def \K{~\rm{K}}

\def \lae{\mathrel{<\kern-1.0em\lower0.9ex\hbox{$\sim$}}}
\def \gae{\mathrel{>\kern-1.0em\lower0.9ex\hbox{$\sim$}}}

\documentstyle[11pt,aasms,tighten,flushrt]{article}
\begin{document}
%\normalsize
\small

\setcounter{page}{1}

\begin{center} \bf 
 ASYMMETRY AND INHOMOGENEITY IN  \\
PROTO AND YOUNG PLANETARY  NEBULAE
\end{center}
%\vspace*{2.0cm}

\begin{center}
Noam Soker\\
Department of Physics, University of Haifa at Oranim\\
%Mathematics-Physics\\
Oranim, Tivon 36006, ISRAEL \\
soker@physics.technion.ac.il 
\end{center}

%\clearpage 

\begin{center}
\bf ABSTRACT
\end{center}
 I study some effects of aspherical mass loss during the last stages
of the asymptotic giant branch (AGB) on the appearance of proto planetary
nebulae (PNs) and young PNs.
 The aspherical mass loss can be small scale inhomogeneities, and/or
axially symmetric mass loss geometry.
 I first examine the role of the dust opacity in the optical band
on the appearance of proto-PNs.
 I conclude that large optical depth will be found in proto-PNs which
are descendant of AGB stars having high equatorial mass loss rates, which
require a stellar binary companion for their existence.
 In these cases light from the central star will reach larger distances
along and near the polar directions, leading to the appearance
of an elongated reflection nebula.
 These proto-PNs will become bipolar-PNs, i.e., PNs with two lobes and
an equatorial waist between them, or extreme elliptical, e.g., a ring,
but no lobes on the two sides of the equatorial plane.
 I then derive the conditions for the enhancement of non-radial
density inhomogeneity by the propagation of the ionization front
at the early PN stages. 
  The ionization will proceed faster in the radial direction
along low density regions.
 The low density regions will be heated earlier, and they 
will expand due to their higher pressures, reducing further their densities.
The opposite occurs in high density regions.
  The condition for this ionization instability to develop is that the
ionization time difference between two direction at the same radius is
longer than the sound crossing time between these two regions.
 This condition for the ionization front instability
can be expressed as a condition on the mass loss rate inhomogeneity,
i.e., its dependence on direction.


%\clearpage

% ======================================================================
\section{INTRODUCTION}
% ======================================================================

 High spatial resolution observations in recent years, mainly with HST,
show that many proto planetary nebulae (PNs) and young PNs possess
complex structures in their inner regions.
 Proto-PNs appear in the optical as reflection nebulae,
with different optical depths along different directions from the central
star to different parts of the nebula (e.g.,
the Egg nebula [CRL 2688], Sahai {\it et al.} 1998a, b; 
more examples in Ueta, Meixner \& Bobrowsky 2000; hereafter UMB).
 In most extreme cases, termed DUPLEX (DUst-Prominent
Longitudinally-EXtended) by UMB, the light from the central star
is completely, or almost completely, blocked in the equatorial plane.
 Many young PNs have filaments, dents, blobs and other inhomogeneities
in their inner regions (e.g., Sahai \& Trauger 1998).

  In the present paper I examine some effects of aspherical mass loss on
the propagation of radiation during the proto-PNs and young PNs stages,
which affect the structures and appearances of the nebulae during
these stages.
 The aspherical mass loss can be an axisymmetrical mass loss, and/or
local inhomogeneities in the mass loss process.
 Such local inhomogeneities may result from instabilities during the proto-PN
stage (e.g., Dwarkadas  \& Balick 1998), or from the mass loss process
itself, as expected from the model of dust formation above magnetic
cool spots (Soker 2000 and references therein).
 In that model, which was proposed to account for the formation
of most elliptical PNs (Soker 2000 and references therein),
the magnetic cool spots on the AGB surface are formed by a weak
magnetic activity, and they facilitate the formation of dust.
 The magnetic field has no dynamic role.
 A weak dynamo activity forms more magnetic cool spots near
the equatorial plane than closer to the poles, thus leading to
axisymmetrical mass loss and the formation of elliptical PNs.
  If the dust formation occurs during the last AGB phase
when mass loss rate is high, the dust shields the region above it
from the stellar radiation (Soker 2000).
 This leads to both further dust formation in the shaded region,
and, due to lower temperature and pressure, the convergence of the
stream toward the shaded region, and the formation of a flow having a
higher density than its surroundings.
 In that model the very large optical depth of the dusty envelope
during the AGB superwind phase increases substantially the departure
from spherical mass loss geometry, both locally and globally.
 This model for axisymmetrical mass loss, which is based on a weak dynamo
activity that leads to the formation of magnetic cool spots,
can explain other features in addition to the axisymmetrical mass
loss itself (Soker \& Clayton 1999).
 Most important, it can operate in very slowly rotating AGB stars,
which can gain the required rotation from a planet companion, or
else be fast rotators on the main sequence.
 Since the departure from spherical mass loss occurs only when opacity near
the AGB surface is large, which means high mass loss rate,
the model explains the higher ellipticity of the inner regions
of many elliptical PNs.
 In extreme cases the inner regions are elliptical while the
halos are spherical (e.g., NGC 6891, Guerrero {\it et al.} 2000). 
 It can also account for local inhomogeneities, such as clumps and filaments,
and it can account for the change in direction of the symmetry axis,
which is later observed as a point symmetric elliptical PN.
 Changes in the direction of the magnetic axis is seen in the Sun
(over a period of $\sim 11 \yrs$), and in Earth (over a period of
hundreds of thousands years). 

 The present paper examines the role of the density inhomogeneities
on radiative transfer at later stages: their role in the optical
appearance of the reflection nebulae during the proto-PN phase ($\S 2$),
and their role in the propagation of the ionization front in young
PNs ($\S 3$).
 It is well established that the ionization front has a substantial
influence on the structures of PNs
(Breitschwerdt \& Kahn 1990; Mellema 1995; Mellema \& Frank 1995;
Chevalier 1997; Sch\"onberner \& Steffen 2000; Soker 1998).
 Here ($\S 3$) I derive the condition for the density inhomogeneities to be
amplified by the ionization front, leading to an ionization front
instability.
 In $\S 4$ I summarize the main results. 

% ====================================================================
\section{POST AGB STARS}              
% ====================================================================

 This section examines the optical radiation from the central star
prior to ionization as it propagates through the dusty nebula.
 For the purpose of this paper it is adequate to take a simple form for
the obscuring AGB wind.
 The mass loss rate per unit solid angle along a specific direction
is $\dot m(\theta) = \dot M (\theta) /(4 \pi)$, and the wind
moves radially at a constant velocity $v_w$.
In general $v_w$ may depend on the direction, but here I simply assume
that it is uniform. 
 I also assume that most of the opacity is due to the intensive mass
loss episode at the termination of the AGB, i.e., the superwind, lasting
for a time $t_{sw}$.
The opacity due to the low density inner regions of the post-AGB
circumstellar matter (e.g., Sch\"onberner \& Steffen 2000)
is neglected. 
 At a time $t$ after the superwind ends the density distribution is
given by
\begin{eqnarray}
\rho(\theta,r)=  \left\{  \matrix{
& {{\dot M(\theta)} / {(4 \pi v_w r^2)}} \qquad &
r_{\rm in} \leq r \leq r_{\rm out} \cr
&  {\rm very} \quad {\rm low} \qquad & 
r< r_{\rm in} \quad {\rm or} \quad r > r_{\rm out} , } \right\}
\end{eqnarray}
where the boundaries of the superwind are
\begin{eqnarray}
r_{\rm in} = v_w t, \quad {\rm and} \quad  
r_{\rm out} =  r_{\rm in} +v_w t_{sw}.
\end{eqnarray}

 The optical depth of the dusty wind is given by
\begin{equation}
\tau_a = \int_{r_{\rm in}}^{r_{\rm out}} \kappa(r) \rho(r) dr=
{{\dot M(\theta) \kappa} \over {4 \pi v_w}}
\left( {{1}\over{r_{\rm in}}} - {{1}\over{r_{\rm out}}} \right)=
{{r_\tau} \over {r_{\rm in}}} 
\left( 1 - {{r_{\rm in}}\over{r_{\rm out}}} \right) =
{{t_\tau} \over {t}} 
\left( 1 - {{t} \over {t + t_{sw}}} \right) 
\end{equation}
where the integration was performed with equation (1) for the density, 
and the length scale $r_\tau$ and time scale $t_\tau$ are defined as 
follows, 
\begin{equation}
r_{\tau} \equiv 
{{\dot M(\theta) \kappa} \over {4 \pi v_w}}
= 5 \times 10^3 
\left( {{\dot M(\theta) } \over {10^{-4} M_\odot \yr^{-1}}} \right)
\left( {{\kappa} \over {150 \cm^2 \g^{-1}}} \right)
\left( {{v_w} \over {10 \km \s^{-1}}} \right)^{-1}
\AU,
\end{equation}
and
\begin{equation}
t_{\tau} \equiv  {{r_\tau}\over{v_w}} 
= 2.4 \times 10^3   
\left( {{\dot M(\theta) } \over {10^{-4} N_\odot \yr^{-1}}} \right)
\left( {{\kappa} \over {150 \cm^2 \g^{-1}}} \right)
\left( {{v_w} \over {10 \km \s^{-1}}} \right)^{-2}
\yrs.
\end{equation}
 The opacity of dust and gas around AGB stars is
$\kappa \sim 20 \cm^2 \g^{-1}$ at $2 \mu$m (e.g., Jura 1986),
and I scaled it to $150 \cm^2 \g^{-1}$ in the optical as it is for the ISM.
 At early stages, $r_{\rm in} \ll r_\tau $, or $t \ll t_\tau$,
the opacity is very large, and no light escapes at all,
as is well known for many upper AGB stars.
 At very late stages $r_{\rm in} \gg r_\tau$, or $t \gg t_\tau$,
and the dusty nebula is transparent.
 The interesting time is when $t \simeq t_\tau $.

 As an example let the mass loss rate (when extended to
a complete sphere) along the equatorial and polar directions be
$\dot M(\pi/2) = 5 \times 10^{-5} M_\odot \yr^{-1}$ and
$\dot M(0) = 10^{-5} M_\odot \yr^{-1}$, respectively.
 The velocity will be taken to be $10 \km \s^{-1}$ in all directions,
and the superwind duration $1000 \yrs$.
 At time $t= 200 \yrs$ the optical depth of the superwind along
the equatorial plane is 5, while it is only $1$ along the polar
direction.
 The central star is substantially attenuated along the equatorial plane,
but only moderately so along the polar directions.
Taking the equatorial flow to be slower will increase the effect further.
 
 The mass loss rate at the end of the superwind phase declines by more
than two orders of magnitude within $\sim 100 \yrs$
(Bl\"ocker 1995; Sch\"onberner \& Steffen 2000), so any effect on
a shorter time scale cannot be accurately treated with
the assumptions made here.
 From equation (5) we learn that for the optical depth to be $\gtrsim 2$,
so that the central star is attenuated by a factor of $\sim 10$,  
for more than $\sim 200 \yrs$ the mass loss should be
$\dot M (\theta) \gtrsim 2 \times 10^{-5} M_\odot \yr^{-1}$.
  This result is interesting since the maximum mass loss rate possible
from radiation momentum transfer is $\dot M_{\rm max} = n_s L /(c v_s)$,
where $L$ is the stellar luminosity, $c$ is the speed of light,
$v_w$ the terminal wind velocity and $n_s$ is the average number of
times a photon is scattered within the outflowing material.
 For most cases $n_s \lesssim 1$ (Knapp 1986). 
 For $L=5000 L_\odot$, $v_w = 10 \km \s^{-1}$, and $n_s=1$
we find $\dot M_{\rm max} = 10^{-5} M_\odot \yr^{-1}$.
 The conclusion is that the dust opacity will obscure almost completely
the central star, at least in some directions, for systems where one or
more of the following occurs:
 (1) luminosity is very high $L \gg 5000 L_\odot$;
 (2) the expansion velocity, $v_w$, (probably in the equatorial
      plane) is very low;
 (3) mass loss rate in the equatorial plane is enhanced by dynamic
     effects of a binary companion.
 A very luminous central star requires a massive progenitor, which if
has a non-spherical mass loss will probably lead to the formation
of a dense equatorial flow, but not necessarily to the formation
of a bipolar PN.
 In any case, a very luminous central star is required, and this is rare.
 Processes (2) and (3) are more likely to occur.
 We note that slowly expanding equatorial flows are found around several
binary systems having orbital separations of $\sim 1 \AU$
(Van Winckel 1999; Van Winckel {\it et al.} 1998; Jura \& Kahane 1999).
 These systems seem to form bipolar PNs (e.g., the Red Rectangle, 
Waters {\it et al.} 1998).
 It seems that a slowly expanding equatorial flow requires binary interaction.
  A high mass loss rate in the equatorial plane due to dynamic effects
can result both from close companions outside the
envelope (e.g., Mastrodemos \& Morris 1999), or from a common envelope
evolution as is evident from the structure of most of the 16 PNs known
to have close-binary nuclei (orbital periods from a few hours to 16 days;
Bond 2000).
 However, in most cases the common envelope evolution will not lead to
the formation of a bipolar-PN but to an elliptical PN
with high equatorial to polar density ratio, e.g.,
ring-like structures but without any lobes (Soker 1997). 
(Bipolar PNs are defined as axially symmetric PNs having
two lobes with an `equatorial' waist between them.)

 The conclusion from this section is that high optical depth,
$\tau \gtrsim 2$, will be found in proto-PNs with high equatorial
mass loss rates, which require their AGB progenitors to
interact with stellar companions.
 The effect of this high optical depth is that light from the central
star penetrates large distances along and near the polar directions,
but not along and near the equatorial plane.
 This leads to the appearance of an elongated reflection nebula
(e.g., UMB).
 However, not all of these will turn into bipolar-PNs.
  Some will form extreme elliptical PNs, e.g., rings without 
lobes on the two sides of the equatorial plane.
 UMB observe 21 reflection proto-PNs with the
Hubble Space Telescope, and classify them into two groups.
 Proto-PNs with highly or completely obscured central stars were termed
DUPLEX (DUst-Prominent Longitudinally-EXtended) reflection nebulae.
 Those with almost no obscuration are termed SOLE
(Star-Obvious Low-level-Elongated) reflection nebula.
The more elongated DUPLEX reflection nebula are mostly a result
of obscuration in the equatorial plane.
No obscuration occurs in the less elongated SOLE reflection nebula.
 I find that out of the 10 DUPLEX proto-PNs presented by UMB
some do not show any signature of lobes, and therefore, I argue, they will
later turn into extreme elliptical PNs rather than bipolar PNs.
I further speculate that these systems may have close binary
nuclei, like the 16 systems listed by Bond (2000).
These are IRAS 19374+2359, and IRAS 23321+6545, and possibly 
IRAS 16342-3814, and IRAS 20028+3910. 

%\clearpage

% ====================================================================
\section{THE EARLY IONIZATION PHASE}              
% ====================================================================

 In this section I derive the condition for the enhancement of density
inhomogeneities during the early PN phase, when the ionization front moves
outward.
 In order to obtain an analytic expression, I neglect the shock preceding
the ionization front (a D-type ionization front), and the motion of
the inner boundary of the superwind inward relative to the rest of
the wind, after it is heated by the ionization front
(e.g., Breitschwerdt \& Kahn 1990; Mellema 1995).
  I also neglect the compression of the inner region by the newly blown
fast wind (Breitschwerdt \& Kahn 1990; Mellema 1995;
Sch\"onberner \& Steffen 2000), and any dependence on longitudes,
and consider only axisymmetrical density profiles. 
 I will therefore use the density profile as given in equation (1).
However, the results are applicable to any dependence
on the direction of the mass loss rate from the progenitor AGB star.
 I make some other simplifying assumptions as indicated during the 
derivation of the condition below.

 The increase in the density inhomogeneities occurs because the ionization
front proceeds with different speeds along directions having different
density profiles.
 Let as assume that along a direction $\theta$ the density is somewhat
higher that along a direction $\theta + \Delta \theta$, because
$\dot M(\theta) > \dot M(\theta + \Delta \theta)$, where the wind velocity is
taken to be constant and equal along the two directions (eq. 1).
 Because of the lower density the ionization front will reach a radius
$r$ along the $\theta + \Delta \theta$ direction earlier than along the
$\theta$ direction.
 Let this time difference be $\Delta t_i(r)$. 
 The ionized region along the $\theta + \Delta \theta$ direction will be
much hotter than the still neutral material along the $\theta$ direction
($\sim 10^4 \K$ compared with $<10^3 \K$), and its thermal pressure 
much higher (assuming that the density ratio is $<10$), so it will 
compress the cooler region along the $\theta$ direction.
 The compression proceeds in the {\it azimuthal} direction,
from $\theta + \Delta \theta$ to $\theta$ with a velocity
$\sim c_s$, where $c_s$ is the adiabatic sound speed.
 The compression continues as long as the ionization front does
not reach the same radius $r$ along the $\theta$ direction,   
and it increases the density in the denser and cooler region,
and decreases the density in the already low density
region along the $\theta + \Delta \theta$ direction.
 I neglect the ionization of the denser region along the $\theta$
direction by the recombination radiation from the matter along the
$\theta + \Delta \theta$ direction. 
 The changes in the densities in the two directions will be significant
{\it only} if the time for the compression wave to cross the distance 
between the two region is shorter than the time difference between the 
ionization times of the two regions $\Delta t_i(r)$.
 The compression time is $\sim r \Delta \theta /c_s$.
 The condition for significant enhancement of the inhomogeneity is
therefore $\Delta t_i(r) > r \Delta \theta /c_s$.
 Taking the limit of small angles and rearranging the equation,
this condition reads
\begin{eqnarray}
t_i^{\prime} \equiv {{d t_i}\over{d \theta}} > {{r}\over{c_s}}.
\end{eqnarray}

 Several characteristics of this instability should be noticed. 
First, this is not the type of instability considered by 
Breitschwerdt \& Kahn (1990), who simply considered the Rayleigh-Taylor 
instability in the interface of the ionized and neutral matter.
 Second, if the solid angle span by the high density region is smaller
than that of the low density region, as in a dense clump,
the instability will result in a dense region extending radially
behind the dense clump.
This case was studied in a previous paper for a few specific cases
relevant to the PN IC 4593 (Soker 1998).
 If it is the low density region that is narrower, it will be ionized first,
and its density will drop further as it expands.
  This will lead to the formation of a faint region within the nebular
shell.
 Third, the instability, when it starts, has a positive feedback (Soker 1998),
due to the increase (decrease) of density of the already denser (tenuous)
region, hence the ionization front will move even slower (faster) along
this direction.
 Fourth, after the denser region is ionized its pressure exceeds that of
its surroundings, and it expands and its density drops. 
 In the present study I only examine the general conditions for the
instability to start, and I do not follow its evolution.
 
 I now turn to express condition (6) in term of the density profile
of the nebula.   
At very early stages the ionizing photon emission rate
$\dot N_\ast$, in photons per second, can be approximated by
a simple linear rise with time.
 Using the post-AGB results presented by Bl\"ocker (1995;
see also Sch\"onberner \& Steffen 2000) and an earlier approximation
(Breitschwerdt \& Kahn 1990; eq. 1 of Soker 1998), I take the
ionization to start a time $t_1$ after the superwind ends, with a
time dependence according to
\begin{eqnarray}
\dot N_\ast (t)  = \dot N_0
\left( {{t-t_1}\over{t_2}} \right), \qquad t>t_1.
\end{eqnarray}
 For the standard case I find from fig. 1 of Breitschwerdt \& Kahn (1990)
$\dot N_0 = 1 \times 10^{47} \s^{-1}$, and $t_2 = 700 \yrs$,
% and as Mellema (1994)
and I take the ionization to star at $t_1= 1000 \yrs$.
 These numbers are very sensitive to the mass of the central star
(e.g., Bl\"ocker 1995), but sufficient to illustrate, and derive the
condition for, the enhancement of initial density inhomogeneities.
 For the density profile I use the same assumption and density profile 
as in the previous section (eq. 1). 

 At the early stages the ionizing flux is low and the ionization front
moves through a dense medium.
 At this stage most ionizing photons go to ionized recombining atoms,
and only a small fraction of the ionizing flux reaches the ionization
front and ionizes new material.
 I therefore approximate the location of the ionization front along
a specific direction by equating the ionizing flux with the recombination
rate per unit solid angle
\begin{eqnarray}
{{\dot N_\ast} \over {4 \pi}}  =
\int _{r_{\rm in}}^{r_f} \alpha n_e n_i r^2 dr,
\end{eqnarray}
where $\alpha$ is the recombination coefficient, and $n_e$ and $n_i$
are the electron and ions number density, respectively.
 To derive the dependence of the ionization front $r_f (\theta)$ on time,
I substitute for the density from equation (1) and for $\dot N_\ast$
from equation (7), and then integrate from $r_{\rm in} = v_w t_i$ to
$r_f (\theta, t_i)$.
 This gives
\begin{eqnarray}
{{r_f (\theta, t_i)} \over {r_{\rm in}}} =
\left( 1 - {{t_i-t_1}\over{t_2}} {{t_i}\over{t_F (\theta)}} \right)^{-1},
\end{eqnarray}
 where I defined the time scale (marked F in eq. 6 of Soker 1998)
\begin{eqnarray}
t_F \equiv
{{\dot M^2(\theta)}\over{4 \pi v_w ^3}}
{{n_e n_i}\over {\rho ^2}} {{\alpha}\over{\dot N_0}} 
\simeq 700 
\left( {{\dot M (\theta)} \over {10^{-5} M_\odot \yr ^{-1}}} \right)^{2} 
\left( {{\dot v_w} \over {10 \km \s^{-1}}} \right)^{-3} 
\left( {{\dot N_0} \over {10^{47} \s^{-1}}} \right)^{-1} 
% \left( {{\alpha}\over{3.5 \times 10^{-13} \cm^3 \s^{-1} }} \right)
\yrs. 
\end{eqnarray}
 More (less) massive cores have a higher (lower) mass loss rate,
but the ionizing photon emission rate is higher (lower) as well,
and $t_2$ shorter (longer).
 These effects may cancel, more or less, their mutual influence
on the product $t_2 t_F$ in equation (9). 
 We also note that the expression for the inner boundary of the superwind
$r_{\rm in} = v_w t_i$ is not accurate, since as this region is ionized its
pressure increases, and the hot material flows inward relative to the
rest of the wind  (e.g., Breitschwerdt \& Kahn 1990; Mellema 1995).
 Therefore, the effective velocity in the relation
$r_{\rm in} = v_w t_i$ may be $<v_w$. 
 Taking $r_f$ from equation (9) for $r$ in the instability 
condition (6) and taking $r_{\rm in} = v_w t_i$ gives for the 
instability condition 
\begin{eqnarray}
{{t_i^{\prime}}\over{t_i}} > 
{{v_w}\over{c_s}}  
\left( 1 - {{t_i-t_1}\over{t_2}} {{t_i}\over{t_F (\theta)}} \right)^{-1}.
\end{eqnarray}

 The next step is to take the derivative of equation (9) with respect
to the angle $\theta$, at a constant value of $r_f$. 
 Rearranging equation (9), taking $r_{\rm in} =v_w t_i$, and then
taking the derivative gives 
\begin{eqnarray}
-r_f {{d}\over{d \theta}} 
\left({{t_i-t_1}\over{t_2}} {{t_i}\over{t_F (\theta)}} \right)
=v_w t_i^{\prime}.
\end{eqnarray}
 In performing the derivation we note that $t_1$ and $t_2$
are constants, while
\begin{eqnarray}
{{d t_F (\theta)}\over {d \theta}} =
{{2}\over{\dot M(\theta)}}
{{d \dot M(\theta)}\over{d \theta}} t_F ,
\end{eqnarray}
so that equation (12) becomes  
\begin{eqnarray}
r_f \left( 
{{t_i-t_1}\over{t_2}} {{t_i}\over{t_F (\theta)}} 
{{2 \dot M^{\prime}}\over{\dot M(\theta)}} 
-{{t_i^{\prime}}\over{t_2}} {{t_i}\over{t_F (\theta)}} 
-{{t_i-t_1}\over{t_2}} {{t_i^{\prime}}\over{t_F (\theta)}} 
\right) =v_w t_i^{\prime}, 
\end{eqnarray}
where $\dot M^{\prime} \equiv d \dot M / d \theta$.
 Substituting $r_f$ from equation (9), and $r_{\rm in} = v_w t_i$ in
equation (14) gives for $t_i^{\prime}$
\begin{eqnarray}
{{t_i^{\prime}}\over{t_i}} = 
{{2 \dot M^{\prime}}\over{\dot M(\theta)}} 
{{(t_i-t_1)t_i}\over{t_2 t_F}}
\left(1+ {{t_i^2}\over{t_2 t_F}}\right)^{-1}.
\end{eqnarray}
 Elimination of $t_i^{\prime}$ from equations (11) and (15) gives 
the instability condition on the mass loss inhomogeneity
\begin{eqnarray}
{{\dot M ^{\prime}(\theta)}\over{\dot M(\theta)}}
> 
{{v_w}\over{c_s}}  
\left(1+ {{t_i^2}\over{t_2 t_F}}\right)
{{1}\over {2 Z (1-Z)}}, 
\qquad {\rm for} \quad t_i>t_1 
\qquad {\rm and} \quad Z>0,
\end{eqnarray}
where  
\begin{eqnarray}
Z(\theta, t_i) \equiv 
{{(t_i-t_1)t_i}\over{t_2 t_F}}.
\end{eqnarray}

 Let us examine the different factors in equation (16). 
The sound speed of the ionized region is $\sim 12-14 \km \s^{-1}$,
depending on its temperature, while the expansion velocity
is $\sim 10-15 \km \s^{-1}$. 
 However, due to the slower motion of the inner boundary of the shell 
at early stages (e.g., Breitschwerdt \& Kahn 1990; Mellema 1995),
its effective value can be somewhat smaller. 
 We therefore can safely take $v_w/c_s \simeq 0.5-1$.
The term $[2 Z(1-Z)]^{-1}$ will reach its minimum value of $2$ 
when $Z=0.5$.  
 For $t_1=1000 \yrs$, $t_2=700 \yrs$, and $t_F = 700 \yrs$, this occurs
at $t_i \sim 1200 \yrs$. 
 By that time the second term on the right hand side of equation (16)
is $1+t_i^2/t_2 t_F  = 3.94$, and the instability condition becomes 
$\dot M ^{\prime}/\dot M \gtrsim 8$. 
The exact minimum value of the r.h.s. of equation (16) (for $v_w=c_s$) 
is $7.82$ and it occurs at $t_i =1184 \yrs$ and $Z=0.445$.
 Taking the mass loss rate to be two (three) times as high as in 
equation (10), with all other parameters being equal, 
so that $t_F = 2800 \yrs$ ($6300 \yrs$), the minimum value 
of the r.h.s. of equation (16) (again, for $v_w=c_s$) is $4.52$, 
($3.82$) and it occurs at $t_i=1534 \yrs$ and $Z=0.418$ 
($1937 \yrs$ and $Z=0.412$). 
 For $t_F \gg t_1$, the time of maximum instability occurs at 
$t_i \gg t_1$. 
Neglecting $t_1$, the r.h.s. of equation (16) reads 
$(v_w/c_s)(1+Z)/[2Z(1-Z)]$. 
 The minimum value of the r.h.s is $2.914 (v_w/c_s)$,
and it occurs at $Z=2^{1/2}-1=0.414$.
 
The third term on the r.h.s.  $[2 Z(1-Z)]^{-1}$  will always reach 
a value as low as $2$ when $Z=0.5$. 
However, when $t_F \ll t_1$ the second term ($1+t_i^2/t_2 t_F$)
will be very large when $Z \simeq 0.5$, and the condition on the
density inhomogeneity will be hard to meet. 
 For example, when the mass loss rate is $\sim 0.5$ ($\sim 0.75$)
of that in equation (10), so that $t_F=200 \yrs$ ($400 \yrs$), 
the minimum value of the r.h.s. of equation
(16) (for $v_w=c_s$) is $18.2$ ($11.0$) and it occurs at 
$t=1062$ and $Z=0.470$ ($t=1115$ and $Z=0.458$).

 The main conclusion from this section is that for most elliptical
PNs, density inhomogeneities as a result of different mass loss rates
along different radial directions during the superwind phase, where
$\dot M \gtrsim 10^{-5} M_\odot \yr^{-1}$,
will be amplified by the propagating ionization front if
$\dot M^{\prime}/\dot M \gtrsim 4$.
 I recall that the derivation of the instability condition applies
to a time-independent mass loss rate, and hence does not apply
to small blobs, but only to inhomogeneities extending to large radial 
distances.
 For example, for the case $t_F=2800 \yrs$, $t_1=1000 \yrs$, and
$t_2=700 \yrs$, we found above that maximum likelihood for the
instability occurs at $t_i \simeq 1500 \yrs$ and $Z=0.418$. 
 From equation (9) we find the ionization front to be at
$r_f = 1.7 r_{\rm in}$, hence the inhomogeneous mass loss rate
lasts for a long time. 
 However, the instability can start much earlier if the density inhomogeneity
is larger: at $t=1100 \yrs$ ($1200 \yrs$) the condition is 
$\dot M^{\prime}/\dot M \gtrsim 15$ ($\gtrsim 8$),
for which $r_f = 1.06 r_{\rm in}$ ($1.14 r_{\rm in}$). 
 If the enhanced, or reduced, mass loss rate spans an angle of 
$0.2=12 ^\circ$, i.e., the density inhomogeneity from the center of the
$12^\circ$ to its edge spans an angle of $\Delta \theta=0.1=6^\circ$,
the instability condition is that the density will be enhanced,
or reduced, by a factor of $e^{15 \Delta \theta} = 4.5$
($e^{8 \Delta \theta} = 2.2$).
 We note that Soker (1998) finds that for a compressed tail to develop
behind a dense clump, the condition is that the ionization front reach 
the clump within $\sim 100 \yrs$ of the beginning of the ionization
($t \lesssim t_1+100 \yrs$), and the density enhancement be
by a factor of $\gtrsim 5$. 
 The present results are compatible with those of Soker (1998).
 Dense clumps can be formed by the mass loss process itself, or from
the interaction process of the fast and slow winds at early stages
(e.g., Dwarkadas  \& Balick 1998).
 Dwarkadas  \& Balick (1998) conduct two-dimensional simulations of winds 
interaction, taking into account the evolution of the fast wind, and find  
that the interaction process is prone to instabilities, which may 
form clumps at early stages. 
 Since the density enhancement is only within a clump, and does not
extend to a large radial distance as with inhomogeneous mass loss
assumed here, the clump density enhancement should be larger,
i.e. $\gtrsim 5$, to form a dense radially extended tail (Soker 1998).

Images of the inner regions of several PNs observed with HST,
e.g., M1-26, He2-142, He2-138, and He2-131, all from Sahai \& Trauger
(1998), reveal many blobs, arcs and filaments with angular width      
of $\sim 0.2-0.5 = 10-30^\circ$. 
 For these inhomogeneities to be amplified by ionization, the density
differences between different regions should have been by a factor
of $\gtrsim 2-5$ as ionization started.
 This is quite plausible in these PNs. 
 Very narrow radially extending structures which span an angle of
$\lesssim 2^\circ = 0.04$, will be amplified even for mass loss rate
enhancement as small as a factor of $1.2-2$.
 
 Finally, I examine the numerical results of Mellema (1995).
 Mellema finds that the ionization front modifies the slow wind density
profile along the azimuthal direction (from pole to equator) in
his model B, but not in his model A.
 In his model B the density profile has a steep variation with the angle
near the pole, where the ionization front modifies the density profile,
while in model A the density variation with the angle is much shallower
in all directions.
 From his density plots I find that within $30^\circ$ from the
symmetry axis (i.e., polar direction) the average value is
$\dot M ^\prime / \dot M \simeq 2.5$, with higher values nearer the
symmetry axis.
 According to the results here, this density inhomogeneity can be amplified,
as indeed happens in the simulation of Mellema (1995), but a shallower
density variation with angle, as in his model A, will not develop
the ionization front instability.

% ====================================================================
\section{SUMMARY}
% ====================================================================

 In the first part of the present paper ($\S 2$) I examined the role of
the dust opacity in the optical band in the appearance of proto-PNs.
 This was motivated in part by the recent HST observations of
proto-PNs by UMB, who classified the proto-PNs into two groups,
SOLE and DUPLEX.
 The conclusion from this study is that a large optical depth,
$\tau \gtrsim 2$, will be found in proto-PNs with high equatorial
mass loss rate.
 The high mass loss rates requires in most cases dynamic effects,
probably from a binary companion.
 Such effects can be gravitational focusing by a binary companion or a
common envelope evolution with a stellar companion.
 In these cases light from the central star will reach larger distances
along and near the polar directions, leading to the appearance
of an elongated reflection nebula.
 The proto-PNs will turn into bipolar-PNs, i.e., PNs with two lobes and
an equatorial waist between them, or will become extreme elliptical,
e.g., a ring, but no lobes on the two sides of the equatorial plane.
 Proto-PNs which will turn into moderate elliptical PNs, i.e., a small
departure from sphericity, will not have high optical depth, and the
light from the central star will not be attenuated much.
 Therefore, while the dust opacity near the stellar surface may lead to
non-spherical mass loss on the upper AGB of the progenitors of
moderate elliptical PNs (Soker 2000), it has no substantial role
thereafter.

 In the second part I examined the conditions for the enhancement of
non-radial density inhomogeneity by the propagation of the ionization front
at early stages.
 The ionization will proceed faster along low density region, which will be
heated earlier than dense regions. 
 The hot low density region will expand due to its higher pressure, and the
density will decrease further (see Mellema 1995 for a specific numerical
simulation).
 The condition for this ionization instability to develop is that the
ionization time difference between two direction at the same radius is
longer than the sound crossing time between these two regions (eq. 6).
 This can be expressed as a condition on the mass loss variation
with the direction $\dot M^{\prime} \equiv d \dot M / d \theta$.
 Assuming a constant mass loss rate with time, and a constant wind
velocity with time and direction, this condition was derived analytically
(eq. 16).
 For typical parameters of elliptical-PNs the ionization
instability will increase density inhomogeneities when
$(\dot M^{\prime}/\dot M) \gtrsim 4 $.
 Therefore, the observed inhomogeneities in young PNs can be larger than
the inhomogeneities of the mass loss process itself.

{\bf ACKNOWLEDGMENTS:} 
 This research was supported in part by a grant from the Israel
Science Foundation, and a grant from the Israel-USA Binational Science
Foundation.

\clearpage

% {\bf FIGURE CAPTIONS} 

%\noindent {\bf Figure 1:}
\end{document}